# A Scalable Heuristic for Fastest-Path Computation on Very Large Road Maps

Renjie Chen  
Max-Planck Institute for Informatics  
Saarbrücken, Germany

Craig Gotsman  
New Jersey Institute of Technology  
Newark, NJ, USA

**Abstract**

Fastest-path queries between two points in a very large road map is an increasingly important primitive in modern transportation and navigation systems, thus very efficient computation of these paths is critical for system performance and throughput. We present a method to compute an effective heuristic for the fastest path travel time between two points on a road map, which can be used to significantly accelerate the classical A* algorithm when computing fastest paths. Our method is based on two hierarchical sets of separators of the map represented by two binary trees. A preprocessing step computes a short vector of values per road junction based on the separator trees, which is then stored with the map and used to efficiently compute the heuristic at the online query stage. We demonstrate experimentally that this method scales well to any map size, providing a better quality heuristic, thus more efficient A* search, for fastest path queries between points at all distances – especially small and medium range – relative to other known heuristics.

## 1. Introduction - Heuristic A* Search

Let $G = <V, E, c>$ be an undirected graph with vertex set $V$, edge set $E \subset V \times V$, and a positive *cost function* on the edges $c: E \to R^+$. The minimal-cost path problem on $G$ is as follows: Given a pair of query points $s, t \in V$, find a path of edges in $G$ from $s$ to $t$, such that the sum of the costs of these edges is minimal among all possible paths. This is called a minimal-cost path and in this way we can generalize the function $c$ to express the minimal cost between *all* pairs of vertices: $c: V \times V \to R^+$.

The subject of computing minimal-cost paths in graphs has been studied for decades and the literature on this topic is vast. Rather than surveying all this here, we refer the interested reader to the recent survey by Bast et al [1]. Our contribution focuses on improving one of the most fundamental approaches to solving this problem – the heuristic A* algorithm.

The A* algorithm [7] is a generalization of the classical (and most basic) Dijkstra algorithm [4], which is notoriously slow. In search of the minimal-cost path from $s$ to $t$, the Dijkstra algorithm traverses many more graph edges and vertices than those on the actual path. In fact, the complexity of computing the minimal cost from $s$ to the single vertex $t$ is essentially the same as computing the minimal cost from $s$ to *all* other vertices in the graph. To wit, the time complexity of the Dijkstra algorithm, even after some optimization [5], is $O(m + n \log n)$, where $m$ is the number of edges and $n$ the number of vertices in the graph. A* improves on this with the help of an *admissible* heuristic function $h: V \times V \to R^+$, such that $\forall s, t: h(s, t) \leq c(s, t)$, namely $h$ is a *lower bound* on $c$. The closer $h$ is to $c$, the faster A* runs, reducing the number of graph edges and vertices traversed during the search for the minimal-cost path. For the trivial $h \equiv 0$, A* reduces to the Dijkstra algorithm, and for the perfectly *informed* $h = c$, A* performs gradient descent on $c$ directly from $s$ to $t$, with no unnecessary search overhead. Ideally, A* should reduce the overhead of the search to some constant multiplier of the number of vertices along the optimal path.

The quest for a good heuristic has also been the subject of much research. One popular approach is to assign to every vertex a vector of "coordinates", in effect embedding in a Euclidean metric space. The heuristic distance between $s$ and $t$ is then the Euclidean distance between the embeddings of the two vertices. Computing an effective embedding is not easy. Given an undirected weighted graph, it is possible to compute an "optimal" embedding in $d$ dimensions for this purpose, by solving a semi-definite program (SDP) [8]. The accuracy of the heuristic will theoretically improve as $d$ increases. Although computing the embedding is done as a preprocessing step on the graph only once, after which the embedding coordinates are stored (requiring $O(dn)$ space complexity) and referred to every time the heuristic is computed, the time complexity of $O(dn^3)$ for this preprocessing is prohibitive for large graphs. The name of the game is, therefore, to optimize the tradeoff between 1) the preprocessing time complexity; 2) the space complexity to store the results of the preprocessing; 3) the time complexity of computing the heuristic based on the stored information when A* runs, and 4) the quality (i.e. accuracy) of the resulting heuristic. We will propose in this paper an admissible heuristic parameterized by a *tree depth $k$*, such that the preprocessing time complexity is $O(2^k n \log n)$, the storage space complexity is $O(kn)$, the heuristic computation time complexity is $O(k)$ and the heuristic quality increases with $k$. In practice, $k$ is taken to be $O(\log n)$, thus the time complexity of preprocessing, storage space complexity of the result, and heuristic computation time complexity become $O(n^2 \log n), O(n \log n), O(\log n)$, respectively. Experimentally we have observed that the overhead of computing the minimal-cost path using this heuristic is a small constant factor of the number of vertices along the path, thus, roughly speaking, the complexity of computing a minimal-cost path in a road map would be $O(\sqrt{n} \log n)$, since the number of vertices along a path in a plane graph is $O(\sqrt{n})$.

## 2. The Global Separator Heuristic (GSH)

Chen and Gotsman [2] proposed a *separator heuristic* for A*, as follows. Let $S \subset V$ be a separator of $G$, namely a set of vertices such that its removal, along with the edges incident on the removed vertices, results in $V$ being partitioned into three sets $U_1$, $S$ and $U_2 = V - U_1 - S$, such that there exist no edges between $U_1$ and $U_2$. This means that $S$ *separates* between $U_1$ and $U_2$ and the separated graph contains at least two connected components, none of them mixing $U_1$ and $U_2$. The properties of graph separators and efficient algorithms to find them has been studied extensively. For two vertices $s, t \in G$, we say that $S$ *separates* $s$ and $t$ if $s \in U_1$ and $t \in U_2$ or vice versa.

The minimal cost function may be generalized to the cost between two *subsets* of vertices: $c: 2^V \times 2^V \to R^+$, by:

$$c(A, B) = \min_{s \in A, t \in B} c(s, t)$$

Let $S$ be a separator of $G$ and $s$ and $t$ be two vertices of $G$. Chen and Gotsman [2] show that the following *separator heuristic* is admissible:

$$h_S(s,t) = \begin{cases} c(s,S) + c(t,S) & \text{if } S \text{ separates } s \text{ and } t \\ |c(s,S) - c(t,S)| & \text{otherwise} \end{cases} \quad (1)$$

Essentially the first case means that if $S$ separates $s$ and $t$, then the minimal cost between $s$ and $t$ is at least the minimal cost from $s$ to $S$ + the minimal cost from $t$ to $S$, since any path from $s$ to $t$ *must* cross $S$. In the second case, that $S$ does not separate $s$ from $t$, a different estimate for $c(s,t)$ can be obtained using the triangle inequality, a generalization of the so-called ALT heuristic [3,6]. The latter is typically much weaker than the former.

Using the separator heuristic based on $S$ means storing the positive value $c(v,S)$ for each $v \in G$. In practice, a number of independent separators $S_i, i = 1,..,k$ are employed, computing and storing in a preprocessing stage *all* the $k$ cost values $c(v,S_i)$ for each vertex $v \in G$ . Each results in a different heuristic value $h_{S_i}(s,t)$. Since all the heuristics are admissible, we can take the final heuristic value to be:

$$h(s,t) = \max_{i=1,\ldots,k} h_{S_i}(s,t) \quad (2)$$

To be effective, the set of separators should cover the map well. This is typically just a set of equally-spaced horizontal and vertical separators. As such, it is a *global* set of separators, earning the heuristic the name Global Separator Heuristic (GSH). While the heuristic is effective, it depends critically on the separation property. If $s$ and $t$ are separated by some $S$, $h_S(s,t)$ will typically be a good estimate of $c(s,t)$, albeit this also depends on the "angle" of the separator relative to $s$ and $t$. If $s$ and $t$ are not separated by any of the $k$ separators, $h(s,t)$ will usually be far from $c(s,t)$. This implies that if the map is very large, a significant number of separators will be required to ensure that most pairs of query points $(s,t)$ are separated by at least one separator, especially if $s$ is close to $t$ (relative to the size of the map), as is typical in navigation applications. Unfortunately, using a large number $k$ of separators could be unpractical, as it would require the storage of a large number of values per graph vertex, which could be prohibitive. Thus it would seem that GSH is not scalable to very large road maps.

### 3. The Local Separator Heuristic (LSH)

We now propose a more sophisticated version of the basic separator heuristic, one that guarantees that any pair of query vertices $(s,t)$, even those close to each other, will be separated with high probability, and the separator angle will not be too bad. This relies on a recursive binary subdivision of the map.

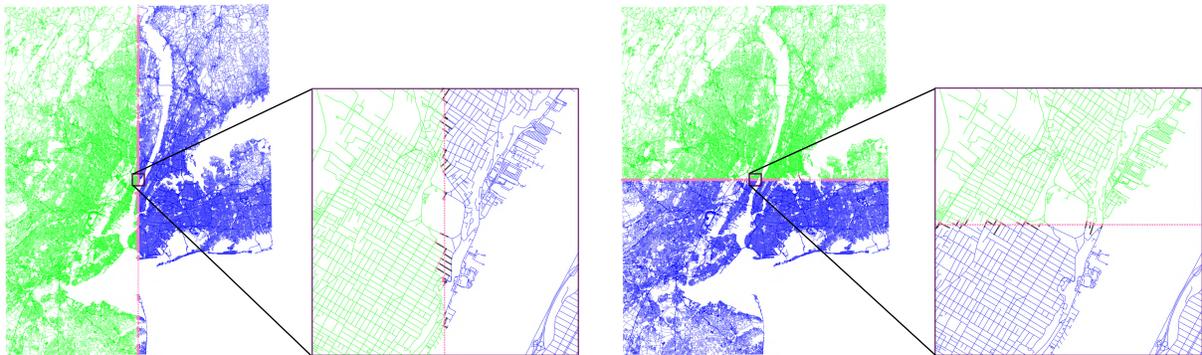

**Figure 1:** Separators of roadmap of NY generated by vertical and horizontal lines. Green vertices are component $U_1$, blue vertices are component $U_2$, and large pink vertices are the separator vertices $S$. The vertices of $S$ are the left (top, resp.) endpoint of the black edges intersected by the vertical (horizontal, resp.) dotted pink line in the plane.

First recall that if the graph is embedded in the plane, namely, each vertex is assigned $(x, y)$ coordinates, and every edge $(u, w)$ drawn as a straight line segment between the position of $u$ and the position of $w$, it is possible to easily generate a separator of the graph by "drawing" a line (or polyline) $L$ on the plane through the graph. $L$ will intersect a subset of the edges $F \subset E$. It is straightforward to verify that the set $S$ defined as the vertices obtained by taking (either) one of the two vertex endpoints of all edges in (a non-empty) $F$ is a separator of $G$. See Fig. 1 for examples

If $G$ is a graph representing a road network, and $B$ its bounding box in the plane, then two independent separators may be generated, as described above. The first, $S^v$, using a vertical line through the center of the box and the second, $S^h$, using a similar horizontal line. Using eq. (1), each of the two separators defines a heuristic, and the final heuristic is obtained, using eq. (2), as the larger of the two values. Since the two separators are perpendicular to each other, they are complementary in "covering" the space of angles. So if the heuristic based on the vertical separator does not do a good job in approximating $c$, there is a good chance that the heuristic based on the horizontal separator will compensate for that. See the example in the top left portion of Fig. 2. The use of these two separators require the storage of the two values $c_1^v(u) = c(u, S^v)$ and $c_1^h(u) = c(u, S^h)$ for each $u \in V$.

If indeed at least one of the two separators $S_1$ or $S_2$ does separate the query vertices $s$ and $t$, then we can finish the heuristic computation here. Alas, with just these two separators, the chances of separating an arbitrary pair of vertices which are close to each other is very small. Thus we must continue to recursively separate the subgraphs generated by the separators and store the relevant minimal costs to these separators. After $k$ levels of recursion, each vertex will have a vector of $k$ costs. We may relate the entries of the cost vectors of two vertices depending on their relative positions. We say that a cost entry of *s is common to* a cost entry of $t$ if they are derived from the same separator. In general, a pair of vertices which are close to each other will have more costs in common than a pair of vertices who are distant from each other. These common costs will be all the costs computed at all the levels higher than the level at which separation occurs. See the example in Fig. 2.

### 3.1 Computing LSH

Assuming $2k$ costs per vertex, the LSH heuristic is computed in two parts: In an offline preprocessing stage, two abstract binary trees of depth $k$ are constructed recursively while generating vertical and horizontal separators. During each recursive process, a $k$-bit binary code and a positive real $k$-vector of costs are assigned to each vertex. One set of codes and costs is computed using the $x$ coordinates of the vertices, and another using the $y$ coordinates. The $x$ coordinates (and similarly the $y$ coordinates) are used by running the pseudo-code in Fig. 3.

In the online query stage, computing the LSH $h(s, t)$ involves comparing the binary codes of $s$ and $t$ from left to right. If the corresponding bits are equal, this indicates a common separator with a common cost that can be used through the triangle inequality. If the corresponding bits are different, this indicates a separator and the separator heuristic may be applied. The procedure terminates when either the codes are exhausted, thus identical, namely no separation is achieved, or when the first nonequal bit is discovered, indicating separation at that level. See the pseudo-code in Fig. 4. The complexity of computing $h(s, t)$ is thus $O(k)$. The chances of separating $s$ and $t$ increases exponentially with depth, so this will happen even when they are quite close to each other, given a reasonable depth. Essentially high probability of separation is achieved for $k = O(\log n)$, where $n$ is the number of vertices in the graph.

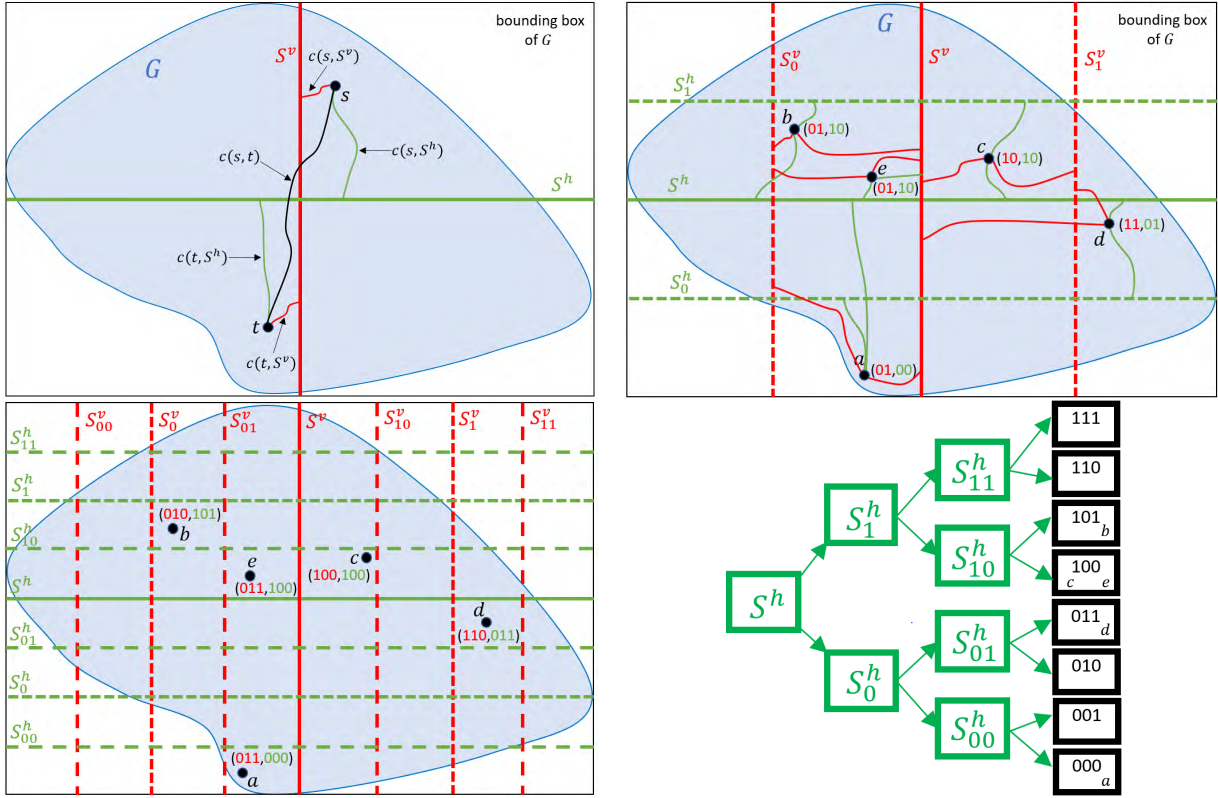

**Figure 2:** The LSH: **(top left)** Top level of the local separator heuristic. The plane embedded graph is separated by a (red) vertical and (green) horizontal line through the middle of the vertical and horizontal extent of the graph. Both these separate $s$ from $t$. The (black) minimal-cost path between $s$ and $t$ has cost $c(s,t)$. The heuristic using $S^v$: $c(s,S^v) + c(t,S^v)$ is much worse than the heuristic using $S^h$: $c(s,S^h) + c(t,S^h)$, which is closer to the true minimal cost $c(s,t)$. Taking $h(s,t) = \max\bigl(c(s,S^v) + c(t,S^v), c(s,S^h) + c(t,S^h)\bigr)$ gives the best of both worlds. **(top right)** Two level separator heuristic. Two new vertical and two new horizontal separators are added to the system, subdividing the two vertical extents and two horizontal extents defined by the top level separator. The binary subscript of the separator indicates in which extent it lies. Any vertex of $G$ can be labeled with a (red) two-bit binary code for the horizontal dimension and a (green) two-bit binary code for the vertical dimension. Derived from the coordinates of the vertex, each code indicates in which of the four quadrants of each dimension the vertex lies. Each vertex has two "costs" in each dimension, measuring the minimal cost from the vertex to the relevant separators. Note that the minimal-cost paths are not necessarily confined to any specific portion of the graph. Since $a$ and $c$ are separated immediately at the top level in both dimensions, then they have only two costs in common, and the heuristic value is determined by only these separators $h(a,c) = \max\bigl(h_{S^v}(a,c), h_{S^h}(a,c)\bigr)$. In contrast, $a$ and $d$ are separated by the top level *vertical* separator, but only by the second level *horizontal* separator, so have one vertical and two horizontal costs in common. Thus $h(a,d) = \max\bigl(h_{S^v}(a,d), h_{S^h}(a,d), h_{S^h_0}(a,d)\bigr)$. Vertices $b$ and $e$ are not separated at all, so have two vertical and two horizontal costs in common: $h(b,e) = \max\bigl(h_{S^v}(b,e), h_{S^v_0}(b,e), h_{S^h}(b,e), h_{S^h_1}(b,e)\bigr)$. The number of costs in common in a given direction is the number of prefix bits they have in common in the respective binary code + 1, capped to the length of the binary code. **(bottom left)** Three level separator heuristic. Each vertex has two three-bit codes. Now $h(a,c)$ is identical to the one-level case, and $h(a,d)$ to the two-level case, but $b$ and $e$ are now

separated also by the vertical $S_{01}^v$ and the horizontal $S_{10}^h$, so $h(b,e) = \max\left(h_{S^v}(b,e), h_{S_0^v}(b,e), h_{S_{01}^v}(b,e), h_{S^h}(b,e), h_{S_1^h}(b,e), h_{S_{10}^h}(b,e)\right)$. **(bottom right)** The binary tree corresponding to the three-level horizontal separator system.

---

**Preprocess**$(G, k)$
for each $v \in G$
      $code(v) := (\ )$;  // empty binary code
      $cost(v) := (\ )$;  // empty cost vector
end
Code$(G, 1, k, x_{min}, x_{max})$;  // $x_{min}$ and $x_{max}$ are the $x$-extents of $G$

---

**Code**$(G, d, d_{max}, x_1, x_2)$  // generate codes and costs for each vertex of $G$ up to depth $d_{max}$
if $d \leq d_{max}$
    $x_c = \frac{(x_1 + x_2)}{2}$;
    $F :=$ edges of $E$ cut by the line $x = x_c$;
    $S :=$ right endpoints of edges in $F$;  // $S$ is separator
    partition $G$ to $(L, S, R)$;
    for each $v \in G$
        $code(v) := \text{concat}(code(v), x(v) \leq x_c \ ? \ 0 : 1)$;
        $cost(v) := \text{concat}(cost(v), c(v, S))$;
    end
    Code$(L, d+1, d_{max}, x_1, x_c)$;
    Code$(R \cup S, d+1, d_{max}, x_c, x_2)$;
end

**Figure 3:** Pseudo-code for the preprocessing stage: constructing binary codes and cost vectors for all vertices of graph $G$ based on $x$-coordinates of the vertices.

---

**heuristic**$(s, t, d)$
$h := 0$;
for $i := 1$ to $d$
   if $code(s, i) == code(t, i)$  // $code(s, i)$ is the $i$'th bit of $code(s)$
      $h := \max(h, |cost(s, i) - cost(t, i)|)$;
   else
      $h := \max\left(h, cost(s, i) + cost(t, i)\right)$;
      return $h$;
   end
   return $h$;
end

**Figure 4:** Pseudo-code of online computation of the local LSH heuristic estimate of the minimal cost between two vertices $s, t \in G$ based on the top $d \leq k$ levels of a separator tree of depth $k$.

## 4. Experimental Results

We have implemented both the Global Separator Heuristic (GSH) of Chen and Gotsman [2] and the Local Separator Heuristic (LSH) described in this paper and compared how informed they are when approximating the fastest travel time on a number of road networks whose edges are weighted with realistic travel times. We used the datasets of Chen and Gotsman [2], who extracted directed graphs on the portions of New York, Colorado and the Bay Area from OpenStreetMap [9]. Table 1 shows the specs of those graphs. We used undirected versions of these graphs, where the edges of the graphs were weighted by the minimal travel time along the two directed edges, which was computed as the Euclidean length of the edge (as computed from the latitude and longitude information per vertex) divided by the maximal speed on that edge, as extracted from OpenStreetMap. In our experiments, we distinguished between pairs of points based on the (Euclidean) distance between them. We randomly selected 3,000 pairs of points in "bins" of distances, e.g.: 1-5 km, 5-10 km, 10-20 km, 20-50 km, 50-100 km, 100-200 km, 200-400 km and 400-750 km. For the state of Colorado, for example, which is approximately a rectangle of size 610×450 km, this covers all possible cases. The $(s,t)$ pairs in each bin were chosen with uniform distribution over area, using the following method for the bin $[a,b]$: $s$ was chosen at random uniformly within the bounding box of the graph, and then "snapped" to the closest map vertex, as long as the snap was not too far. $t$ was then chosen also at random within an annular region centered at $s$ with inner radius $a$ and outer radius $b$ and snapped to the closest vertex as long as the snap was not too far and $||s-t|| \in [a,b]$. We then compared the true fastest travel time $c(s,t)$ with the heuristic $h(s,t)$, when varying the number of levels of the binary tree, thus the number of costs used (and stored) per vertex, between 1 and 9 (in each of the two dimensions).

| Roadmap | Colorado (COL) | Bay Area (BAY) | New York (NY) |
| --- | --- | --- | --- |
| **Physical dimensions (km)** | 623 x 450 | 178 x 225 | 85 x 112 |
| **Vertices** | 5,154,659 | 3,092,249 | 1,579,003 |
| **Edges** | 5,400,186 | 3,351,919 | 1,744,284 |

For each pair of vertices $(s,t)$, we measure the relative *quality* of the heuristic:

$$\text{qual}(s,t) = \frac{h(s,t)}{c(s,t)}$$

which is a value in [0,1] reflecting how accurate the heuristic is. The closer this number is to 1, the more informed the heuristic is. The quality of the heuristic is the mean of this quantity over all possible pairs $(s,t)$.

The *efficiency* of a heuristic in conjunction with A* is measured as the number of vertices on the fastest path divided by the total number of vertices traversed by A*:

$$\text{eff}(s,t) = \frac{\#vertices(\text{fastest\_path}(s,t))}{\#vertices(\text{A*\_traversal}(s,t))}$$

The closer this number is to 1 – the more efficient the heuristic is. The efficiency of the heuristic is the mean of this quantity over all possible pairs $(s,t)$. The best possible efficiency on a road network is typically 40%-50%, since any variant of A* must traverse at least the fastest path vertices and also their immediate neighbors. When an uninformed heuristic is used, the efficiency can sometimes drop dramatically to the vicinity of 1%, meaning 100 vertices of the graph are explored for every one vertex along the fastest path.

Figs. 5-6 show the qualities and efficiencies of LSH vs. GSH obtained in the experiments we performed on the roadmaps of Colorado, New York and the Bay Area, one graph per distance bin, as a function of separator tree depth. The results for Colorado have been plotted in two separate graphs to avoid clutter. Each bin is colored separately, the dashed line for GSH and the solid line for LSH. We compare the performance of a global system of separators using $2k$ separators ($k$ in each dimension), thus $2k$ costs per vertex, with that of the local system of separators of depth $k$ (one system in each dimension), thus also $2k$ costs per vertex. The global set of $2k$ separators was obtained by uniformly partitioning the map bounding box into $k+1$ vertical and horizontal strips. It is straightforward to see that the quality curves must be monotonically increasing in the local case, but not necessarily in the global case.

For the Colorado map, LSH ultimately obtains a quality of 87%-94% for all distances, requiring more depth, as expected, for shorter distances. while GSH obtains 73%-83% on the short distances (< 50km) and 90%-95% on long distances (50km – 750km). The most significant difference, obviously, is in the short distances, where, for example, in the 1-5 km range, LSH achieves a quality of 94% at depth 9 compared to a quality of 77% for GSH. For the 5-10 km range, LSH achieves a quality of 90% at depth 7 compared to 72% for GSH.

The difference in efficiency is also most significant at shorter distances. For distances of 1-5 km, LSH achieves 75% compared to 60% for GSH and at distances of 5-10 km, LSH achieves 63% compared to 42% for GSH.

Similar results are obtained for the maps of BAY and NY. Since these span areas smaller than COL, the first bin covers a smaller range of distances: 1-2 km, for which a separator tree of depth 9 was needed. The last bin was 100-200 km for BAY and 50-100 km for NY. For 1-2 km, LSH provided on BAY a quality of 92% and efficiency of 68%, vs. 75% and 53% for GSH. Similarly, on NY, LSH provided a quality of 90% and efficiency of 52%, vs. 78% and 38% for GSH.

Looking back at the heuristic computation code in Fig. 4, for which there are two different ways to compute the heuristic, depending on whether separation is achieved or not, it is interesting to determine whether separation is indeed achieved, and if so, whether the final value of the heuristic is actually determined by that separator. Fig. 7 provides answers to these questions about the role played by separation. For each map and distance range, we plot (in the dashed line) the probability that a separator was found between $s$ and $t$, and (in the solid line) the probability that that separator determined the final heuristic value (i.e. was larger than any other value considered). The results show that with enough depth, separation is always achieved, albeit at shallow depths for longer distances, as expected. A little surprisingly, the separator does *not* always determine the final heuristic value, going from 45%-55% on short distances, increasing to close to 100% on the longer distances.

To visualize better the effect of LSH vs. GSH, and appreciate how an increase in depth can dramatically increase the efficiency of LSH, which cannot be expected for GSH, Fig. 8 shows a number of fastest paths between vertices on our three different maps. The source vertex $s$ is marked in black and the target $t$ in magenta and the fastest path between them in blue. Red marks edges and vertices traversed by A* in computing the path using the GSH heuristic, and green marks the same when the GSH heuristic is used. The cyan edges in the background are those traversed by the simple Dijkstra algorithm, which is equivalent to A* using the completely uninformed heuristic $h \equiv 0$. The figure shows the improvement in efficiency when a tree of depth 7 is used compared to a tree of depth 3 (or, equivalently, the number of separators used for GSH is increased from $2 \times 3 = 6$ to $2 \times 7 = 14$). For example, on the COL map, going from 3 to 7 improves the efficiency from 9.7% to 33.6% for LSH, compared to a minor improvement from 9.4% to 9.7% for GSH. On the BAY map, an improvement from 14.9% to 53.1% for LSH, compared to the much smaller improvement from 12.2% to 15% for GSH. On NY, LSH improves from 12.2% to 46.6% vs. 9.0% to 12.3% for GSH. For some of these examples, even more dramatic improvements are possible for deeper separator trees (depths 8 and 9).

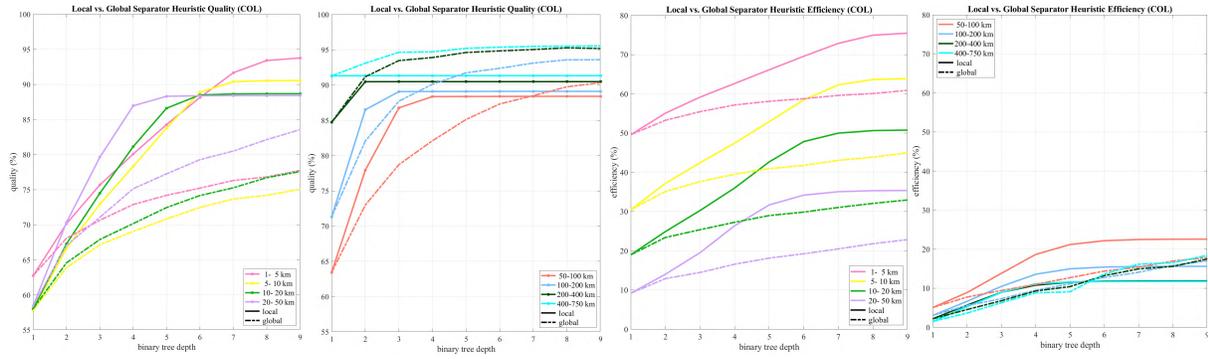

**Figure 5:** Heuristic qualities and efficiencies on the road map of Colorado for $(s,t)$ pairs of different distances, as a function of the separator tree depth. For the local separator heuristic (LSH), marked by solid curves, quality monotonically increases with tree depth. For the global separator heuristic (GSH), marked by dashed curves, quality tends to increase with the number of separators, but at a much lower pace. Large distances require shallower trees than short distances, which benefit much more from tree depth.

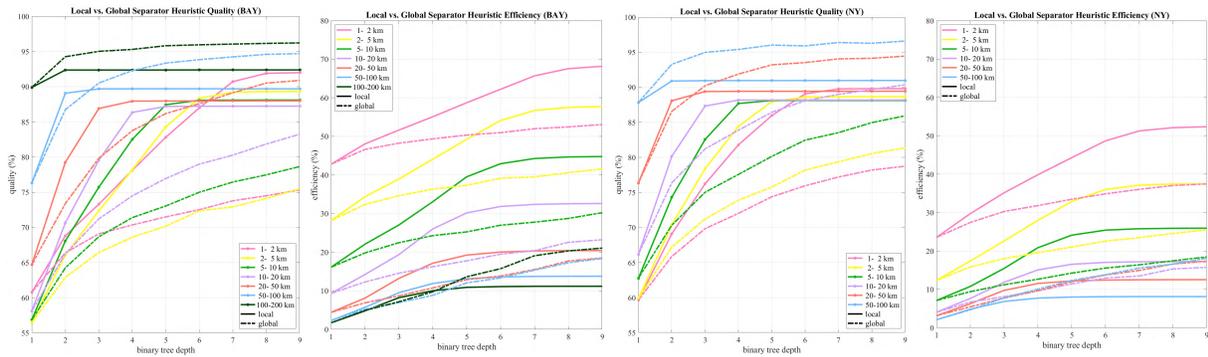

**Figure 6:** Heuristic qualities and efficiencies on the road maps of BAY and NY for $(s,t)$ pairs of different distances, as a function of the separator tree depth.

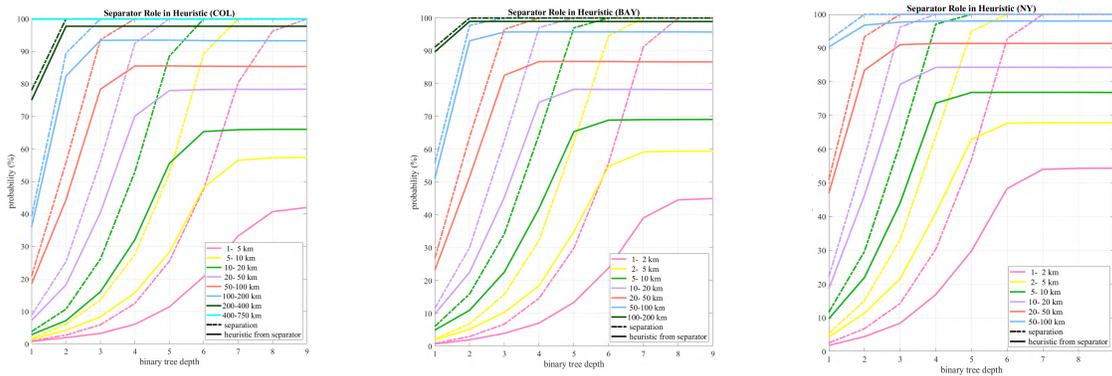

**Figure 7:** The role played by a separator in determining the heuristic. The solid curves are the probability that a separator exists between $s$ and $t$ in the tree. The dashed curves are the probability that a separator exists *and* the final heuristic value is determined by that separator, as opposed to one of the other graph separators (that do not separate $s$ and $t$, thus rely only on the triangle inequality).

5. **Summary and Discussion**

We have described a separator-based admissible heuristic for A*, which is based on a binary tree of separators, thus scales well to provide an informed estimate of the fastest travel time between two vertices in a road map, at short and long distances. The shorter distances are captured well by the deeper levels of the tree. We have shown that this approach, which we call the Local Separator Heuristic (LSH), easily outperforms the simpler approach, called Global Separator Heuristic (GSH), where the map is simply partitioned in both dimensions by a fixed number of separators, simply because GSH does not scale well. Loosely speaking, to be effective, the required number of costs stored per vertex is $O(n)$ for GSH vs $O(\log n)$ for LSH, where $n$ is the number of vertices in the graph.

Our LSH heuristic has been formulated in this paper and experimented with for undirected graphs, but is immediately applicable also to the more realistic case of directed graphs (especially when the graph models a road network). As shown by Chen and Gotsman [2], dealing with directed graphs merely doubles the number of costs stored per graph vertex. They also show that the quality and efficiency of separator-based heuristics is similar for both types of graphs.

Our description of LSH dictates the storage of a binary code vector along with the vector of costs per vertex. While this is convenient, making for a very efficient computation of the heuristic in an online query (as described in Fig. 4), it is not absolutely necessary, as the code may be easily reproduced online during the heuristic computation by a binary search-type procedure on the vertex coordinates.

Our simple implementation of LSH using a binary subdivision procedure of the map bounding box, namely, at each recursive step the bounding box is partitioned into two along the center of the box. While this is the natural subdivision method, in some applications it may be advantageous to subdivide the space differently. An obvious alternative would be to partition along the median of the graph vertex count, namely at the point where half of the graph vertices would lie on either side of the partition. For graphs with non-uniform distribution of vertices over the map, this may be quite different from the center of the box. We have experimented with this type of subdivision procedure as well as a number of other more sophisticated "adaptive" subdivision procedures leading to trees of non-uniform depth, but have concluded that the difference in their performance is insignificant. Thus we see no reason to deviate from the basic approach we have described.

One aspect of our work that could be improved is the time complexity of the preprocessing stage. Currently it is $O(n^2 \log n)$, which is quite slow, and we speculate that it could still be optimized down to $O(n \log n)$. This is especially important for dynamic traffic maps, in which the edge costs (i.e. travel times) change frequently, forcing the preprocessing to be repeated. Along the same note, it would be interesting to devise an efficient *update* procedure for the vertex cost vectors in the event of a few isolated changes to the edge weights.

Finally, we should mention that we have run our experiments on modest road maps representing single states of the USA, containing a few million vertices. In the largest map – Colorado – a binary tree of depth 9 sufficed to achieve very good results. Since the tree depth scales logarithmically with the number of vertices in the map, a simple calculation shows that a tree of depth 12 would suffice to deal well with a map of the entire continental USA.

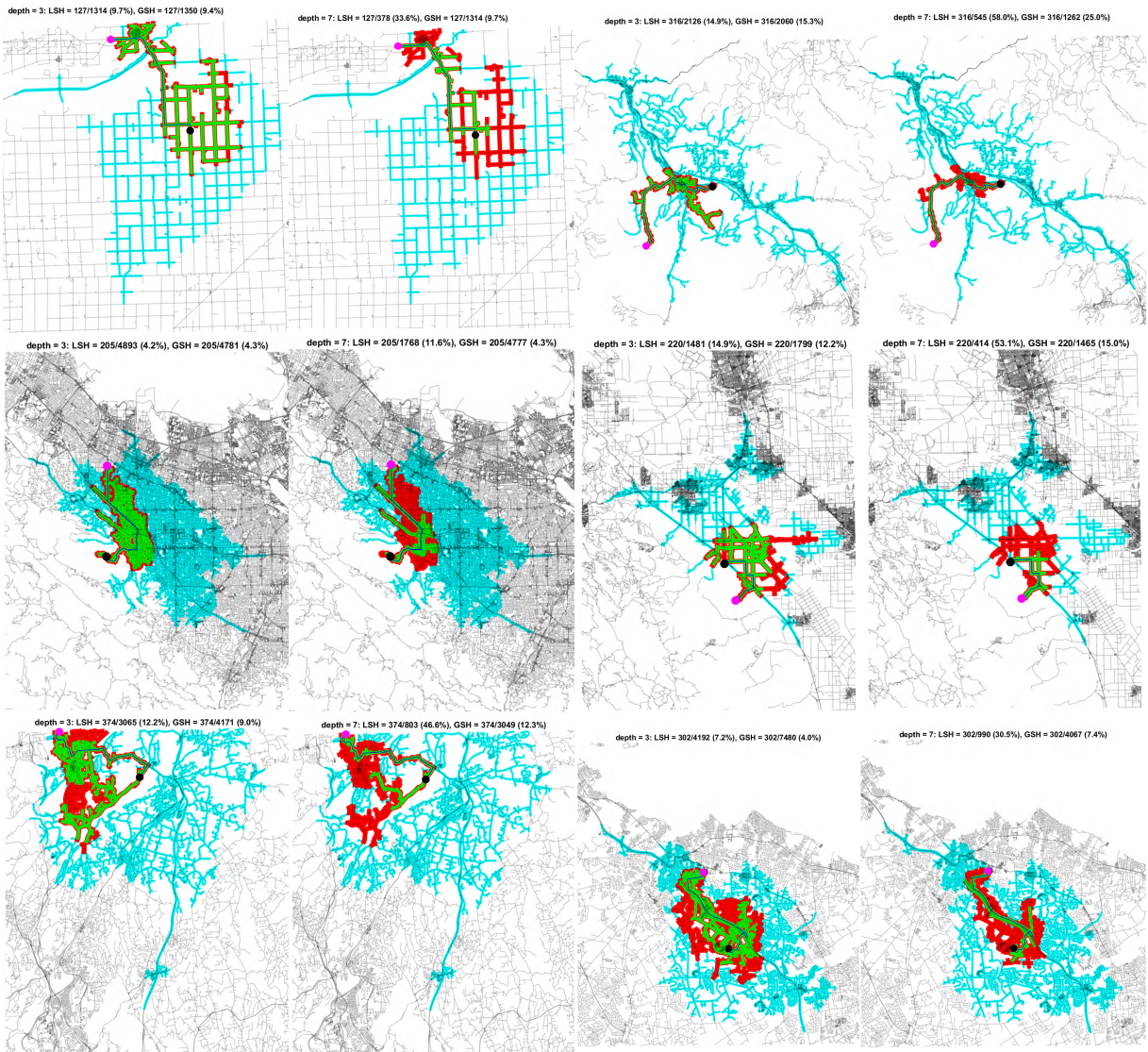

**Figure 8:** Heuristic efficiencies of sample A* runs between black source vertex and magenta target vertex using different heuristics and 6 or 14 costs/vertex (equivalent to two trees of depth 3 or 7). Colored vertices (and edges) are those traversed by A*: Blue – no heuristic, Red – GSH, Green – LSH. Blue path is resulting fastest path. Heuristic efficiencies are in brackets. **(top)** COL **(middle)** BAY **(bottom)** NY. Note the significant difference in efficiency between the two heuristics at the higher depth of 7.